\documentclass[aps,preprint,noshowpacs,superscriptaddress,groupedaddress]{revtex4}  

\usepackage{graphpap}
\usepackage[dvips]{graphicx}
\usepackage[dvips]{graphics}
\usepackage{color}
\usepackage{hyperref}
\usepackage{amsmath}
\usepackage[normalem]{ulem}

\graphicspath{{./figures/}}
\usepackage{SIunits}

\usepackage{xcolor}

\usepackage{nomencl}
\makenomenclature

\begin{document}

\title{Real-time measurement of nanotube resonator fluctuations in an electron microscope }

\author{I.~Tsioutsios}
\thanks{These authors contributed equally}
\author{A.~Tavernarakis}
\thanks{These authors contributed equally}
\author{J.~Osmond}
\affiliation{ICFO-Institut de Ciencies Fotoniques, The Barcelona Institute of Science and Technology, 08860 Castelldefels
(Barcelona), Spain}
\author{P.~Verlot}
\thanks{Corresponding author: pierre.verlot@univ-lyon1.fr}
\affiliation{ICFO-Institut de Ciencies Fotoniques, The Barcelona Institute of Science and Technology, 08860 Castelldefels
(Barcelona), Spain}
\affiliation{Universit{\'e} Claude Bernard Lyon 1, UCBL, Domaine Scientifique de La Doua, 69622 Villeurbanne, France}
\author{A.~Bachtold}
\affiliation{ICFO-Institut de Ciencies Fotoniques, The Barcelona Institute of Science and Technology, 08860 Castelldefels
(Barcelona), Spain}

\date{\today}


\begin{abstract}
Mechanical resonators based on low-dimensional materials provide a unique platform for exploring a broad range of physical phenomena. The mechanical vibrational states are indeed extremely sensitive to charges, spins, photons, and adsorbed masses. However, the roadblock is often the readout of the resonator, since the detection of the vibrational states becomes increasingly difficult for smaller resonators. Here, we report an unprecedentedly sensitive method to detect nanotube resonators with effective masses in the $10^{-20}\,\mathrm{kg}$ range. We use the beam of an electron microscope to resolve the mechanical fluctuations of a nanotube in real-time for the first time. We obtain full access to the thermally-driven Brownian motion of the resonator, both in space and time domains. Our results establish the viability of carbon nanotube resonator technology at room temperature and pave the way towards the observation of novel thermodynamics regimes and quantum effects in nano-mechanics.

\end{abstract}

\maketitle 
Mechanical resonators based on nanotubes~\cite{Lassagne2009,Steele2009,Benyamini2014}, nanowires~\cite{Arcizet2016,Rossi2016}, graphene~\cite{Bockrath2014,DeAlba2016,Mathew2016}, and semiconductor monolayers~\cite{Lee2013,Morell2016} have attracted considerable interest. Upon decreasing their size, mechanical resonators become increasingly sensitive to adsorbed mass~\cite{GilSantos2010,chaste2012,Hanay2012} and external forces~\cite{Moser2014}. These systems hold promise for exploring a broad range of physical phenomena, such as magnetic resonance imaging~\cite{Degen2009,Nichol2013}, surface science~\cite{Tavernarakis2014}, out-of-equilibrium thermodynamics~\cite{Gieseler2014}, and light-matter interaction~\cite{Gloppe2014}. However, the efficient motion detection of these small resonators remains a challenging task, despite intense efforts in improving detection methods over the last decade~\cite{Sazonova2004,Weber2014,Singh2014,Bunch2007,Stapfner2013,Garcia2007}. This has prevented the study of some of the most fundamental properties of these systems, such as the time evolution of the Brownian motion and other types of displacement fluctuations of nanotube resonators~\cite{barnard2012fluctuation,zhang2014interplay,Micchi2015,koh2015thermally,Sun2016,Maillet2016}.

The displacement fluctuations of a resonator with a high quality-factor are fully characterized by recording the time-evolution of the two quadratures of motion. The displacement $x$ is given by 
\begin{equation}
    \label{eq:brownian}
    x(t) = X_1(t)\cos(\Omega_0t)+X_2(t)\sin(\Omega_0t)
\end{equation}
with $X_1$ and $X_2$ the two quadratures, $\Omega_0/2\pi$ the mechanical resonance frequency, and $t$ the time. The quadratures are obtain via real-time demodulation of the motion signal, which must resolve the vibrations at a rate much faster than the mechanical resonance frequency. This condition has never been achieved with resonators based on nanotubes, graphene, and semiconductor monolayers, due to insufficient sensitivity of the tranduction schemes employed thus far. These include methods based on electrical detection~\cite{Sazonova2004,Weber2014,Singh2014}, optical interferometry~\cite{Bunch2007,Stapfner2013}, and scanning probe microscopy~\cite{Garcia2007}. The relatively poor efficiency of these techniques  is essentially explained by the weak interaction overlap between the measurement probe (e.g. electric or optical fields) and the nanometer scale mechanical resonator. With the advantage of much reduced interaction volumes, electron microscopy has also been considered for the study of the fluctuating behaviors in nanomechanical resonators~\cite{Treacy1996}; where long-term imaging is used for quantitatively characterizing the variance of thermally induced mechanical motion. More recently, electron microscopy has been utilized for ultra-sensitive detection and manipulation of resonant properties of nanomechanical resonators with picogram effective masses~\cite{Nigues2015}.

In this work, we push this novel e-beam nanoelectromechanical approach for detecting the thermally-induced fluctuations of attogram-scale ($10^{-21}\,\mathrm{kg}$) carbon nanotube resonators in real-time for the first time. We focus the electron beam of a scanning electron microscope (SEM) at a fixed position, and measure the intensity of the electrons scattered inelastically from the nanotube. This allows high signal-to-noise detection of thermally driven resonances at frequencies up to $10\,\mathrm{MHz}$. This also enables the full access to the real-time trajectories of the quadratures. We demonstrate that nanotube resonators undergo thermally-driven Brownian motion. That is, the motion is driven by thermal noise, with negligible electron beam backaction noise. This new method also allows us to simultaneously measure the time-evolution of two modes polarised perpendicularly. The interaction between the electron beam and the nanotube may be increased by depositing amorphous carbon onto the nanotube. This results in increased energy absorption which may yield to dynamical backaction effects and notably to self-oscillating behaviours that are observed in the present work. We demonstrate that the electron beam of a SEM is a unique tool for addressing the nature of the mechanical motion.

\textbf{Motion detection with a focused electron beam.} 
Our detection scheme relies on coupling nanomechanical motion to a focused beam of electrons \cite{Nigues2015,Buks2000}. Electron beams can be focused to spot sizes approaching the diameter of nanotube resonators, ensuring a much higher interaction overlap compared to usual capacitive or optical techniques used to detect nanomechanical motion \cite{Sazonova2004,Weber2014,Singh2014,Bunch2007,Stapfner2013}. The principle of the detection works as follows \cite{Nigues2015}: The collisions between the electron beam and the nanotube yield to the emission of so-called secondary electrons (SEs), which result from inelastic scattering mechanisms. The displacements of the nanotube within the electron beam create a strong modulation of the secondary electrons current, whose fluctuations are detected by means of a high bandwidth scintillator. Note that previous measurements of nanotubes using electron beams did not resolve neither the power spectrum nor the real-time evolution of their mechanical fluctuations \cite{Treacy1996}. The principle of the experiment is depicted on Fig. 1(a). The samples are mounted onto a 3-dimensional positioning stage hosted in a commercial SEM delivering a highly focused, ultra-low noise electron beam \cite{Nigues2015}. Importantly, the SEM chamber is thoroughly pumped in order to secure a high vacuum level, preventing any significant electron-beam assisted spurious deposition mechanism on the nanotube~\cite{ding2005mechanics}. The mechanical resonators discussed below consist of single-clamped nanotubes that are anchored at the edge of silicon wafers (Fig. 1 (b,c)).  We specifically consider three distinct devices, labelled D$1$ to D$3$. This detection method can also be employed with doubly-clamped nanotubes, but the results are not shown here. 

We first operate the SEM in the conventional "scanning mode", with the electron beam being scanned over the surface of the sample and the SEs response simultaneously acquired. Figure 2(a) shows a typical SEs image obtained by scanning a suspended carbon nanoresonator representative of those investigated in this work (  D1). The image seems increasingly blurred towards the upper end of the nanoresonator, which is interpreted as a consequence of position noise \cite{Treacy1996}. When the latter is large compared to the spatial extension of the electron beam, the integrated current $\mathcal{I}(\mathbf{r_{\mathrm{p}}},\Delta t)=\frac{1}{\Delta t}\int_0^{\Delta t}\mathrm{d}tI(\mathbf{r}_{\mathrm{p}},t)$ becomes simply proportional to the probability $P(\mathbf{r}_{\mathrm{p}},\Delta t)$  to find the object at the electron beam position $\mathbf{r}_{\mathrm{p}}$ within the integration time $\Delta t$ (here $I$ denotes the SEs emission rate). Provided that the image integration time is long with respect to the motion coherence time, the signal becomes proportional to the asymptotic probability, that is the spatial Probability Density Function (PDF) associated with the position noise \cite{wang1945theory}, $P_{\infty}(\mathbf{r}_{\mathrm{p}})=\lim\limits_{\Delta t\rightarrow+\infty} P(\mathbf{r}_{\mathrm{p}},\Delta t)$. For a singly clamped, unidimensional Euler-Bernoulli beam vibrating in the scanning plane and at thermal equilibrium, this probability is given by:

\begin{eqnarray}
P_{\infty}(\mathbf{r}_{\mathrm{p}})&=&u(\mathbf{r}_{\mathrm{p}}.\mathbf{e_{\mathrm{CNT}}})\frac{1}{\sqrt{2\pi}\sigma_{\mathrm{th}}}e^{-\frac{(\mathbf{r}_{\mathrm{p}}.\mathbf{e_1})^2}{2\sigma_{\mathrm{th}}^2}}\label{eq:1},
\end{eqnarray}

with $\mathbf{e_{\mathrm{CNT}}}$ and $\mathbf{e_1}$ respectively denoting the axis and vibrational direction of the carbon nanotube resonator, $u$ its fundamental mode shape \cite{Krishnan1998} and $\sigma_{\mathrm{th}}^2$ the thermal motion variance (the origin of the referential being taken at the anchor point of the resonator). Figure 2(b) shows two cross-sections of Fig. 2(a) (black, dashed arrow) obtained with two distinct scanning rates, corresponding to short and long electron beam exposure, respectively. The cross-sections confirm the Gaussian scaling of the SEs emission rate, and enables to extract the same value of the thermal motion variance $(\sigma_{\mathrm{th}})^2\simeq(14\,\mathrm{nm})^2$, independent from the exposure duration, showing that the nanomechanical dynamics is negligibly affected by the electron beam as further discussed below. Measurements on other nanotube cantilevers show that $\sigma_{\mathrm{th}}$ remains constant upon rotating the nanotube along its axis. This is because nanotube cantilevers feature two fundamental modes polarized perpendicularly with similar effective masses and similar resonant frequencies, so that the variance of the projected thermal motion is independent of the rotation angle. 

To further establish the vibrational origin of this motion imprecision, we turn the SEM into "spot mode", where the electron beam is fixed at a given position. We set the electron beam at the tip of the resonator and acquire the SEs current fluctuations using a spectrum analyser (\cite{Nigues2015,Buks2000}). Figure 2(c) shows two peaks centred at $\Omega_0/2\pi\simeq 5.58\,\mathrm{MHz}$ and $6.33\,\mathrm{MHz}$, consistent with the expected resonant behaviour. The signal to noise ratio of the low-frequency resonance is 14~dB, limited by broadband background scattering (linear spectral decay, arising from the detector cutoff frequency of the SEM). We verified that these peaks vanish when decoupling the electron beam from the nanoresonator, confirming their motional origin.

The measurements in Figs. 2(b,c) enable to determine the basic mechanical properties of the carbon nanotube resonator. The motion variance can be written as a function of the lateral spring constant $k$, $\sigma_{\mathrm{th}}^2=k_BT/k$, yielding $k\simeq 2.1\times 10^{-5}\,\mathrm{N}\mathrm{m}^{-1}$, with the temperature $T=300\,\mathrm{K}$. On the other hand, the effective mass $m_{\mathrm{eff}}$ expresses in terms of the spring constant and mechanical resonance frequency, $m_{\mathrm{eff}}=k/\Omega_0^2\simeq 17\,\mathrm{ag}$. From the values of the lateral spring constant and nanotube length, the radius can be evaluated on the order of $r\simeq 2\,\mathrm{nm}$, assuming that the nanotube contains one wall (Supplementary Information). Using the length and radius of the nanotube and the mass density of pristine graphene, we obtain that $m_{\mathrm{eff}}\simeq 2.8\,\mathrm{ag}$. The difference between this value and the mass measured above is attributed to a thin layer of contamination adsorbed on the nanotube.

As mentioned earlier, the above measurements are weakly sensitive to e-beam induced dynamical effects, which would result in strong distortions of the motion PDF:  Because of the much reduced surface of the electron beam compared to the motion variance, the average electrical power received by the nanotube resonator strongly depends on the e-beam position (see Supplementary Information). To verify the impact of the electron beam on the nanomechanical dynamics, we subsequently image the nanotube resonator by setting the e-beam dwell time $t_d$ to short and extended values ($t_d\ll 2\pi/\Gamma_0$ and $t_d\gg 2\pi/\Gamma_0$, Fig. 2(b) left and right, respectively). Short dwell times enable to decrease the effective electrical power received by the nanotube, at the expense of decreased signal-to-noise ratio, whereas extended dwell times correspond to higher e-beam effective exposure. The obtained motion PDF remains Gaussian and with very similar variances in both cases, excluding any significant dynamical contribution arising from the electron beam fluctuations. Additionally, the electron beam may be responsible for asymmetric electrothermal dynamical backaction effects, which either cool or amplify the nanomechanical vibrations depending on the side from which the nanotube is exposed \cite{Nigues2015}. However, Fig. 2(b) shows that the motion PDF remains symmetric at lower dwell time, which indicates that dynamical backaction effects remain negligible.

\textbf{Damping rate of nanotube resonators.}
The potential of nanomechanical devices relies on their ultra-sensitive dynamical behaviour, which requires the ability to operate them close to their fundamental limits and in real-time \cite{Albrecht1991,cleland2002noise}. To do so, we connect the scintillator output of the SEM to an ultra-fast lock-in amplifier, which we use for demodulating the quadratures of the electromechanical signal around the mechanical resonance frequency. Figure 3(a) shows the fluctuations spectrum of the out-of-phase quadrature obtained with device D2, with the demodulation frequency being set to $356\,\mathrm{kHz}$. Two peaks are observed, with comparable widths and heights.Using the measured resonant frequency and the spatial PDF measurement, we obtain $\sigma_{\mathrm{th}}^2\simeq (31\,\mathrm{nm})^2$ and $k\simeq 4.8\times 10^{-6}\,\mathrm{N}\mathrm{m}^{-1}$. Figures 3(b) and 3(c) further show the spectrum of the electromechanical signal as demodulated around each resonance frequency. The data adjust very well to Lorentzian models (plain lines), suggesting that the nanotube resonator behaves as a linearly damped, 2-dimensional harmonic oscillator.

To address the origin of the observed mechanical linewidths, we compute the autocovariance of the energy of the electromechanical signal $\mathcal{C}_{I}(t,t+\tau)=\langle (I^2(t+\tau)-\langle I^2\rangle)(I^2(t)-\langle I^2\rangle)\rangle$, with $\tau$ the measurement delay time, and $\langle ...\rangle$ statistical average. The energy autocovariance has indeed the property to be insensitive towards frequency noise (see Supplementary Information), enabling the pure extraction of the mechanical damping rates, with the additional benefit of minimal driving amplitude, therefore avoiding possible nonlinear artefacts \cite{stipe2001noncontact}. For a linear, stationary driven non-degenerate 2-dimensional mechanical oscillator, this energy autocovariance is independent of $t$ and can be shown to read as (see Supplementary Information):
\begin{eqnarray}
\mathcal{C}_{I^2}(\tau)&=&g_x^4\sigma_x^4 e^{-\Gamma_x\tau}+g_y^4\sigma_y^4 e^{-\Gamma_y\tau}\nonumber\\
&+&2g_x^2g_y2^2\sigma_x^2\sigma_y^2 e^{-\frac{\Gamma_x+\Gamma_y}{2}\tau}\cos{\Delta\Omega\tau},\label{eq:2}
\end{eqnarray}
with $\Delta\Omega/2\pi$ the frequency splitting between the two modes and with $\Gamma_p$, $\sigma_p^2=\langle p^2\rangle$ and $g_p=\frac{1}{\sqrt{2}}\frac{\partial I}{\partial p}$ the mechanical damping rate, the motion variance and the electromechanical coupling rate associated with each vibrational direction ($p\in\{x,y\}$), respectively. The terms on the first line of Eq. \ref{eq:2} identify to the individual energy components associated with each mode, whereas the second line simply corresponds to the acoustic beat between the two motional polarizations.  Figure 3(d) shows the electromechanical energy autocovariance corresponding to the spectrum shown on Fig. 3(a). The experimental data (dots) are found to adjust very well to the theoretical model set by Eq. \ref{eq:2} (plain line).

It is interesting to compare the "apparent" quality factors $\tilde{Q}_p=\Omega_p/\delta\Omega_p$ obtained from the fits of the quadrature spectrum ($\delta\Omega_p$ denoting the mechanical linewidth associated with each vibrational direction, $p\in\{x,y\}$), to the "intrinsic" quality factors  $Q_p=\Omega_p/\Gamma_p$, measured via the autocovariance of the energy. The measurements presented on Figs. 3(a) and 3(d) are consecutively repeated a number of times and used for extracting the corresponding damping parameters, yielding to $\tilde{Q}_x=412\pm 89$, $\tilde{Q}_y=570\pm 123$, $Q_x=583\pm 70$ and $Q_y=583\pm 50$. These values show no significant difference between the apparent and intrinsic quality factors, which establishes that the measured decoherence is dominated by dissipation mechanisms in the carbon nanotube resonator. In other words, the Duffing restoring force and the mode-mode coupling forces, which arise from inertial nonlinear effects in singly-clamped beams \cite{villanueva2013nonlinearity}, are weak enough so that motional fluctuations do not induce sizable dephasing~\cite{barnard2012fluctuation,zhang2014interplay}.

\textbf{Motion statistics of nanotube resonators.}
We now turn our attention to the statistical analysis of nanomechanical motion. We insist that this aspect is indispensable for resolving the nature and origin of the vibrational state: Indeed, fundamental differences in vibrations, such as those reported in Refs. \cite{Gloppe2014,barnard2012fluctuation,rugar1991mechanical,teufel2011sideband}, can be resolved only by measuring their motion quadrature distribution \cite{richter2002nonclassicality}. Figure 4(a-i) (resp. 4(b-i)) shows the time evolution of the   motion quadratures $(X_{1}(t),X_{2}(t))$ of $x$ (resp. $(Y_{1}(t),Y_{2}(t))$ of $y$), defined as the cross-phase, slowly varying components of mechanical motion,  $x(t)=X_{1}(t)\cos{\Omega_x t}+X_{2}(t)\sin{\Omega_x t}$ and $y(t)=Y_{1}(t)\cos{\Omega_y t}+Y_{2}(t)\sin{\Omega_y t}$. The corresponding real-time displacements $x(t)$ and $y(t)$ are shown on Figures 4(a-ii) and 4(b-ii). Figures 4(a-iii) and 4(b-iii)) show the quadratures cross-correlation functions $C_X(\tau)=\langle X_{1}(t)X_{2}(t+\tau)\rangle$  and $C_Y(\tau)=\langle Y_{1}(t)Y_{2}(t+\tau)\rangle$ associated with each trajectory. These correlations are found to vanish below the $10\%$ level and can therefore be safely neglected. Figures 4(a-(iv,v)) and  4(b-(iv,v)) show the histogram of the normalized motion quadratures, which are all found to be Gaussian distributed with unit variance (plain lines). In total,  these measurements show that the quadratures of the nanomechanical fluctuations in each vibrational direction describe a Brownian motion \cite{wang1945theory}, consistent with a 2-dimensional mechanical resonator at thermal equilibrium. These results establish that nonlinear mechanical effects in singly-clamped nanotube resonators at room temperature remain weak. Figure 4(c) and 4(d) further show the corresponding motion trajectory and associated histogram in real-space, confirming a bivariate, symmetric Normal distribution of the position noise.

To complete our study, we evaluate the spatial correlations defined as $2\sigma_x\sigma_yC_{xy}(\tau)=\langle \{X_{1}(t)+iX_{2}(t)\}\{Y_{1}(t+\tau)-iY_{2}(t+\tau)\} \rangle$. The result is reported on Fig. 4(e), where the real and imaginary parts are shown separately. The very low level of correlations indicates that potential landscape nonlinearities have negligible effects, to first order \cite{Gloppe2014}. Finally, we note that the 2-dimensional, non-degenerate nature of suspended nano-cantilevers provides them with the peculiar property to develop short-term spatial correlations under random external driving, such as the one resulting from measurement backaction. These correlations manifest through strong distortions in the electromechanical spectrum  \cite{caniard2007observation}, which are not observed in our measurements (see Fig. 3(a) and Supplementary Information). This indicates the absence of any random external driving source, and in particular confirms the innocuity of the electron beam towards the vibrational state. 

\textbf{Self-oscillation of nanotube resonators.} The above presented results have been obtained under optimized experimental conditions, with e-beam induced dynamical effects being kept negligible (see Fig. 2). In particular, it is essential to maintain excellent vacuum conditions in order to avoid unwanted contamination processes, which are enhanced under e-beam exposure (\cite{ding2005mechanics}, see also Supp. Inf.]. The deposited material (e.g. amorphous carbon) may indeed act as an efficient energy absorber, yielding non-instantaneous heating of the CNT resonator. An important aspect of our work relies in the fact that we are able to straightforwardly address the dynamical consequences of such effects. 

It has been previously shown that electro/opto-mechanical coupling can change the effective temperature of the resonator in a cavity-free scheme \cite{Nigues2015,Tavernarakis2016}. The carbon nanotube motion evolves in a delayed force gradient \cite{metzger2004cavity} leading to electromechanically-induced dynamical effects which may alter the mechanical behavior. Figure 5(a) shows two mechanical spectra  acquired with the e-beam being positioned at two distinct locations of an amorphous carbon “contamination island” grown at the edge of the resonator (device D3). Efficient cooling (broad curve) and heating (narrow curve) of a carbon nanotube resonator are consequently observed.

A mechanical resonator undergoing ponderomotive heating is susceptible to enter the instable regime of self-oscillation~\cite{kippenberg2005analysis}. In this case, the mere mechanical spectrum may not allow distinguishing between a stable and an instable regime, due notably to the unavoidable presence of frequency noise. Here, we show that by extracting in real-time the motion quadratures, we are able to unveil signatures in the phase-space trajectory indicating a self-oscillation.

Figure 5(b) shows a hole in the PDF suggesting oscillations of the associated motion quadratures. In an instable regime the nanotube is undergoing self-sustained oscillations rather than the ordinary thermal random walk. The phase-space trajectory is then confined in a well-defined region with a non-zero mean amplitude value. The demodulation frequency detuning and the residual frequency noise lead to the exploration of all four phase-space quadrants. Similarly, Figure 5(c-(i,ii)) depict the histograms of the motion quadratures, both presenting a non-Gaussian distribution as it would be expected for a non-thermal state.

\textbf{Discussion.} The present work demonstrates that our novel measurement method enables to detect the vibrations of nanotube-based resonators with masses as low as $17\,\mathrm{ag}$. The measurement of such ultra-low mass resonators raises the question of the limits of our approach. Besides the strong, sub-nanometre confinement of the electron probe, the other key element of our scheme lies in the layout of the device. The absence of any electron scatterer within the immediate vicinity of the free-standing nano-object enables a very high SEs contrast, which is at the origin of the high motion sensitivity (see Supplementary Information).

On a more fundamental side the measurement is responsible for a random backaction that may affect the vibrational fluctuations of the measured objects \cite{braginsky1995quantum}. While such effects are not observed in the present work (where the investigated devices are driven by thermal forces ranging between $(2\,\mathrm{aN}\mathrm{Hz}^{-\frac{1}{2}})^2$ and $(10\,\mathrm{aN}\mathrm{Hz}^{-\frac{1}{2}})^2$), they may become significant for nanotube resonators with higher quality factors. Indeed, e-beam quantum backaction acting on thick semiconducting scatterers has recently been evaluated to be on the order of $(1\,\mathrm{aN}\mathrm{Hz}^{-\frac{1}{2}})^2$ under standard operating conditions \cite{Nigues2015}, which should be in reach e.g. at low temperature, where the mechanical quality factors are found to be enhanced by several orders of magnitude \cite{Moser2014}. Though certainly representing a limit from the sensing point of view, this points out that singly clamped nanotube resonators are devices of choice for probing and controlling quantum properties of electronic beams.

Lastly, we would like to once more attract the attention on a very important and useful characteristic of our singly clamped suspended nanotube resonators, that is their 2-dimensional vibrational nature. This property makes these resonators sensitive to spatially induced motion correlations, resulting in strong distortions in their electromechanical spectrum \cite{Tavernarakis2016}.  These signatures (such as the non-Lorentzian resonance lineshapes in response to an external piezo drive, see Supplementary Information) enable to address the presence and nature of external driving forces, with no further calibration being required. In particular, it is interesting to note that these nanomechanical objects are expected to surpass the limits set by quantum backaction in principle \cite{caniard2007observation}, which has so far never been observed and would represent an important step from the perspective of Quantum Measurement. This 2-dimensional behaviour has also been highlighted as a strong asset in the context of ultra-sensitive nanomechanical detection, related to the corresponding ability to self-discriminate the external noise mode in phase-coherent measurements \cite{GilSantos2010,gavartin2013stabilization}, which will be highly beneficial to our systems.

\textbf{Conclusion}
We have shown that the focused electron-beam of a SEM operated in spot mode allows to detect the noise dynamics of attogram-scale singly clamped suspended carbon nanotubes resonators in real-time. We have demonstrated that a SEM operated in spot mode behaves as a stereoscope with our devices, enabling the tri-dimensional reconstruction of their motion fluctuations in real-time. We have presented a detailed analysis of the 2-dimensional noise trajectories both in space and time, and shown that such small objects behave as Brownian particles evolving in a two-dimensional harmonic potential. Our work paves the way towards the exploration of novel thermodynamic regimes at scales  which have been so far inaccessible experimentally.
\section*{Methods}
\textbf{Sample fabrication}
The nanotubes used in this work are grown via chemical vapor deposition on silicon substrates. Nanotubes are attached to the surface of the substrate by van der Waals forces. Some of the nanotubes extend beyond the substrate edges, thus forming singly-clamped resonators (Fig. 1(b-c)), with lengths in the $100\,\mathrm{nm}-10\,\mu\mathrm{m}$ range.
\section*{Acknowledgements}
We thank A. Reserbat-Plantey for discussions related to sample fabrication. We acknowledge the ERC starting grant 279278 (CarbonNEMS), the EE Graphene Flagship (contact no. 604391), the Foundation Cellex, Severo Ochoa (SEV-2015-0522) of MINECO, the Fondo Europeo de Desarrollo Regional (FEDER), the grant MAT2012-31338 of MINECO, and the Generalitat (AGAUR) for financial support. P.V acknowledges support from the French National Funding Agency for Reasearch (ANR) (NOFX2015 and QDOT).
\section*{Author Contributions}
I.T has fabricated the samples, the process being developed by A.B. I.T, A.T, J.O and P.V have performed the measurements. I.T, A.T and J.O have contributed to the development of the experimental setup. P.V, I.T and A.T have performed the data analysis, with inputs from A.B. P.V has performed all theoretical calculations. P.V has written the supplementary information, with contributions from I.T, A.T and A.B. A.T and P.V have proposed and designed the project. P.V and A.B have supervised every steps of the work. All authors have discussed the results. P.V  and A.B have written the manuscript, with contributions from I.T and A.T.

\section*{References}
\bibliographystyle{Nature}

\clearpage

\begin{figure}
\centering
\includegraphics[width=\columnwidth]{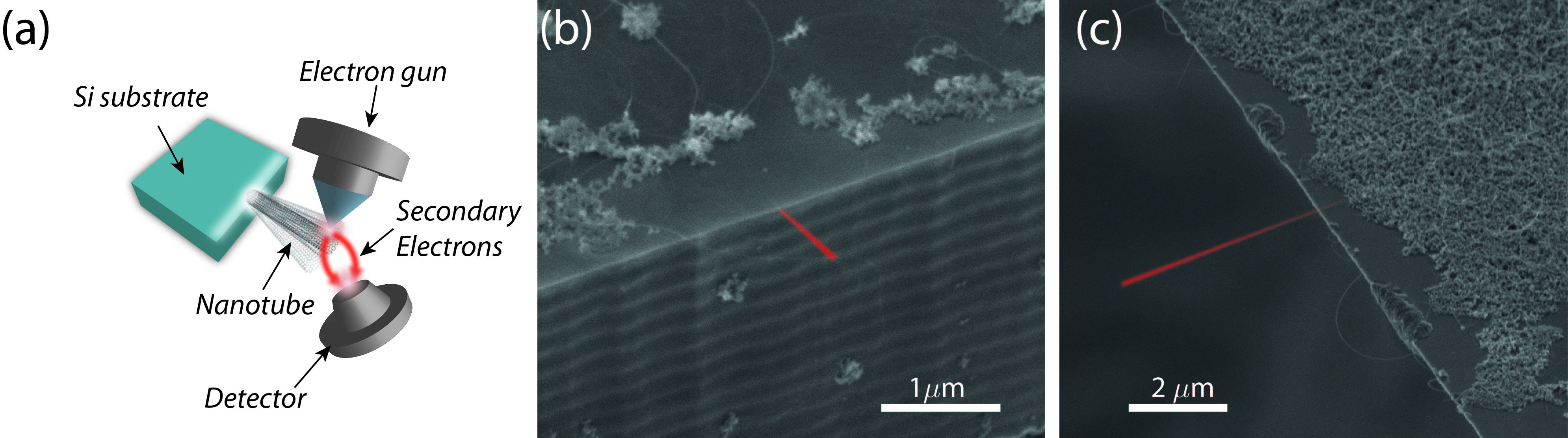}
\caption{\textbf{Experimental setup and systems.} (a) Schematic of the experimental setup. The carbon nanotube resonators are mounted inside a Scanning Electron Microscope (SEM), where their motion is detected via the Secondary Electrons (SEs) emission \cite{Buks2000,Nigues2015}, whose fluctuations are collected at the video output of the SEM and further sent to a spectrum analyser. (b-c) $2$ SEM micrographs showing typical singly-clamped carbon nanotube obtained with our Chemical Vapour Deposition (CVD) growing method. The catalyst (white flakes) is spread on the Silicon substrate (darker parts) which is further CVD processed, resulting in the growth of ultra-low diameter carbon nanotubes, some of which being found to be singly clamped (highlighted in red, false colors).} 
\label{fig:figure1}
\end{figure}

\begin{figure*}
	\centering
	\includegraphics[width=\columnwidth]{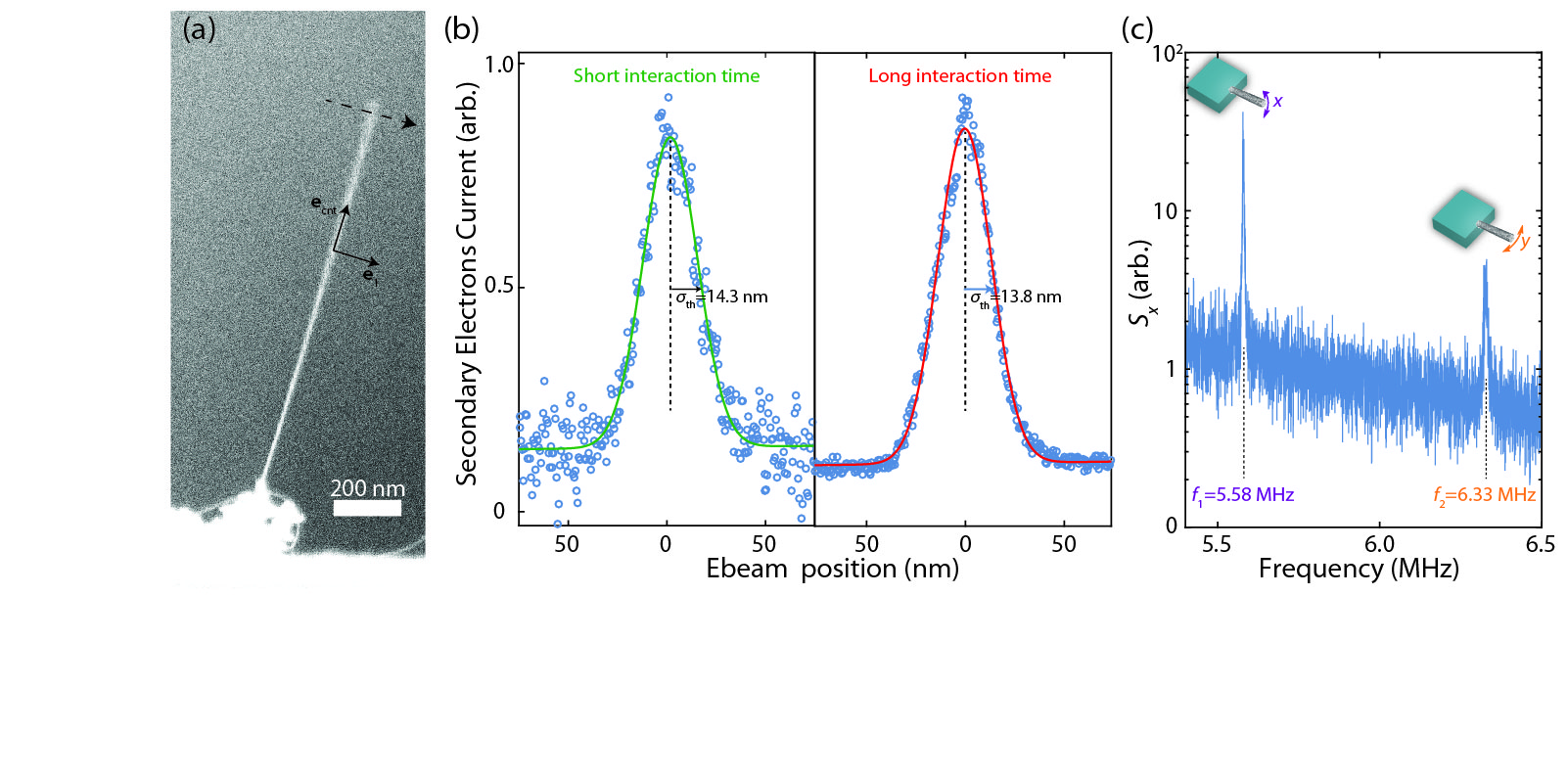}
	\caption{\textbf{Basic mechanical characterization using a SEM.} (a) Magnified SEM micrograph showing a suspended carbon nanotube representative of those used in the present work (device D1). The image is clearly blurred towards the upper end of the nanotube, characteristic of the thermal excitation of its fundamental vibrational mode. (b) Intensity profile taken across the section denoted by the dashed arrow on Fig. 2(a), obtained with fast and slow scanning rate (left and right, respectively). The straight line corresponds to a Gaussian fit, enabling to extract the motion variance $\sigma_{\mathrm{th}}^2\simeq(14\,\mathrm{nm})^2$. (c) Power spectral density of the electromechanical signal. The SEM is operated in spot mode, the electron beam being set at the edge of the carbon nanotube. The resulting SEs fluctuations are collected at the SEM video output and further sent to a spectrum analyser. Two peak are observed around the fundamental resonance frequency $\Omega_x/2\pi=5.58\,\mathrm{MHz}$ and $\Omega_y/2\pi=6.33\,\mathrm{MHz}$, corresponding to the two perpendicular directions of vibration of the nanotube resonator.}
	\label{fig:figure2}
\end{figure*}

\begin{figure}
	\centering
	\includegraphics[width=\columnwidth]{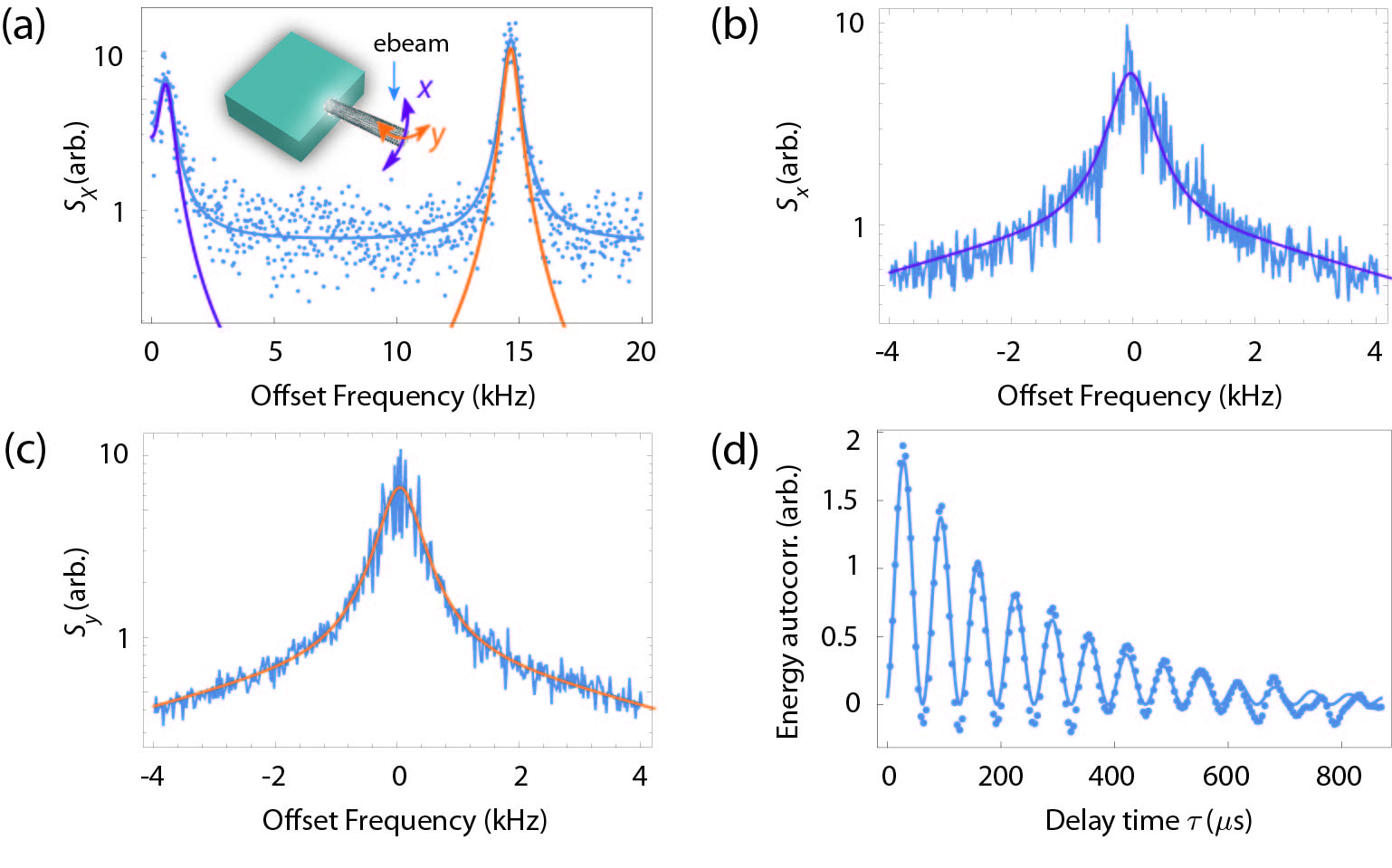}
	\caption{\textbf{Real-time dynamics of a carbon nanotube resonator (device D2)}. (a) Motion quadrature spectrum of a free running carbon nanotube resonator. The quadratures of the electromechanical signal are demodulated at $\Omega_{\mathrm{d}}/2\pi=356\,\mathrm{kHz}$ using an ultra-fast lock-in amplifier while the electron beam being set at the upper edge of the device. The spectrum is obtain as the Fourier transform of the $1\,\mathrm{s}$-averaged autocorrelation of the out-of-phase electromechanical quadrature. Two peaks are observed, associated to the motion imprecision in each vibrational direction of the resonator. Straight lines correspond to Lorentzian adjustments (individual in purple and orange, dual incoherent sum in blue), enabling to extract both mechanical resonance frequencies $\Omega_x/2\pi=356.577\,\mathrm{kHz}$ and $\Omega_y/2\pi=370.243\,\mathrm{kHz}$ and the values of the apparent quality factors $\tilde{Q}_x=\Omega_x/\delta\Omega_x=541$ and $\tilde{Q}_y=\Omega_y/\delta\Omega_y=591$. (b) Motion spectrum associated with $x(t)$. The data are obtained by demodulating the electromechanical signal around frequency $\Omega_x/2\pi$ and further computing the Fourier transform of its $1\,\mathrm{s}$-averaged autocorrelation. The straight line corresponds to a single Lorentzian fit with additional, incoherent background. (c) Same as (b) for $y(t)$. (d) Electromechanical energy autocovariance calculated as $\mathcal{C}_{I^2}(t,t+\tau)=\langle (I^2(t+\tau)-\langle I^2\rangle)(I^2(t)-\langle I^2\rangle)\rangle$, with $I$ the SEs current, $t$ the time, $\tau$ the measurement delay time, and $\langle ...\rangle$ statistical average. The straight line stands for the theoretical adjustment set by Eq. \ref{eq:2}, yielding to the values of the intrinsic quality factors $Q_x=582$ and  $Q_y=559$.}
	\label{fig:figure3}
\end{figure}

\begin{figure}
	\centering
	\includegraphics[width=\columnwidth]{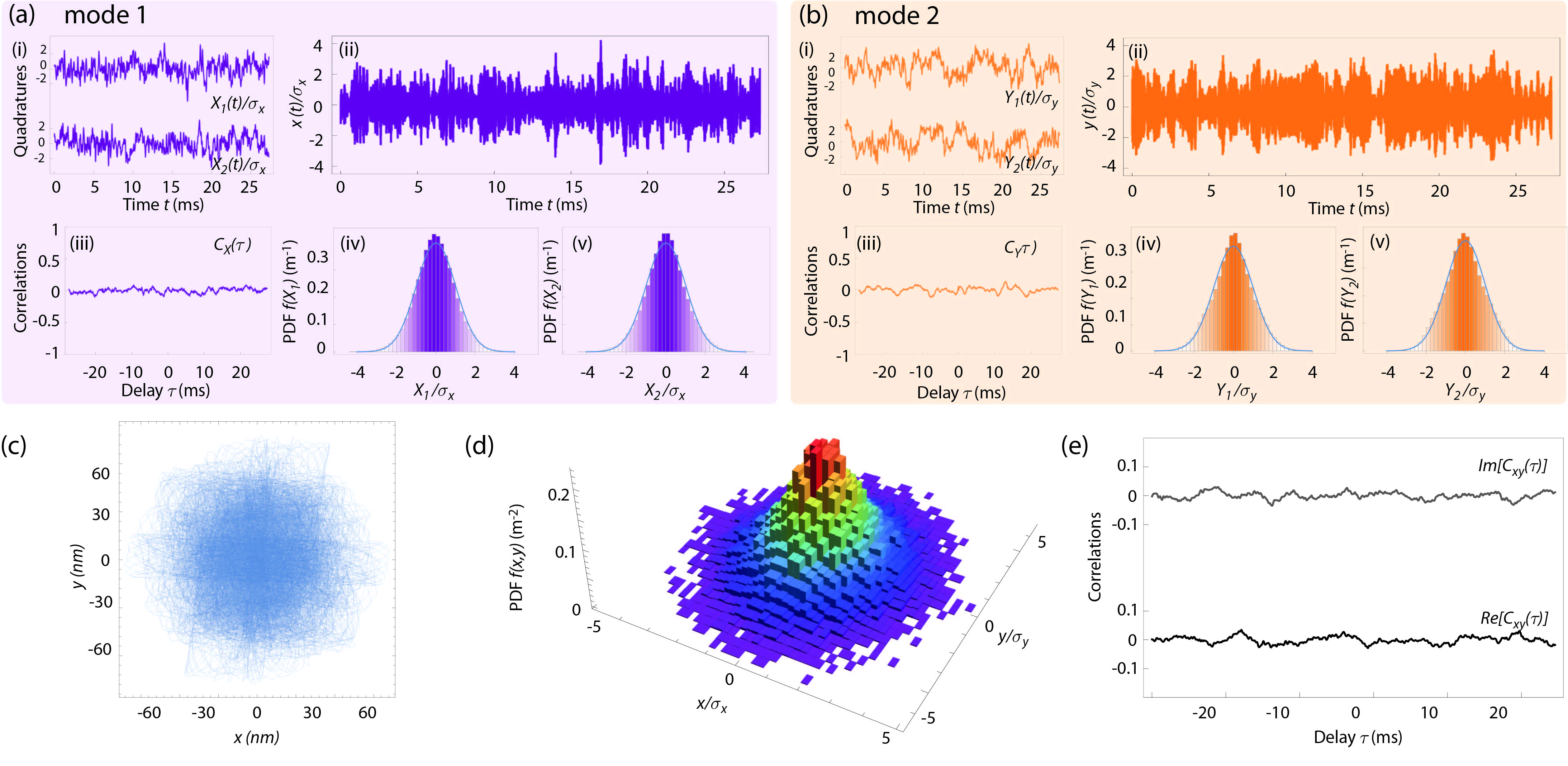}
	\caption{\textbf{Motion statistics of a carbon nanotube resonator (device D2).} (a) (a-i) Time evolution of the motion quadratures for mode $1$. (a-ii) Corresponding real-time evolution of the position $x(t)$. (a-iii) Quadratures cross-correlation for mode $1$. (a-iv,v) Histograms of the normalized quadratures associated with $x(t)$. Straight lines are Gaussian curves with unit variance. (b) Same as (a) for mode $2$. (c) Nanomechanical trajectory  $(x(t),y(t))$ in real-space. (d) Histogram of the nanomechanical trajectory $(x(t),y(t))$ in real-space. (e) Spatial correlations as a function of time. The upper and lower curves stand for the imaginary and real parts of the spatial correlation function, respectively (see text).}
	\label{fig:figure4}
\end{figure}

\begin{figure}
	\centering
	\includegraphics[width=\columnwidth]{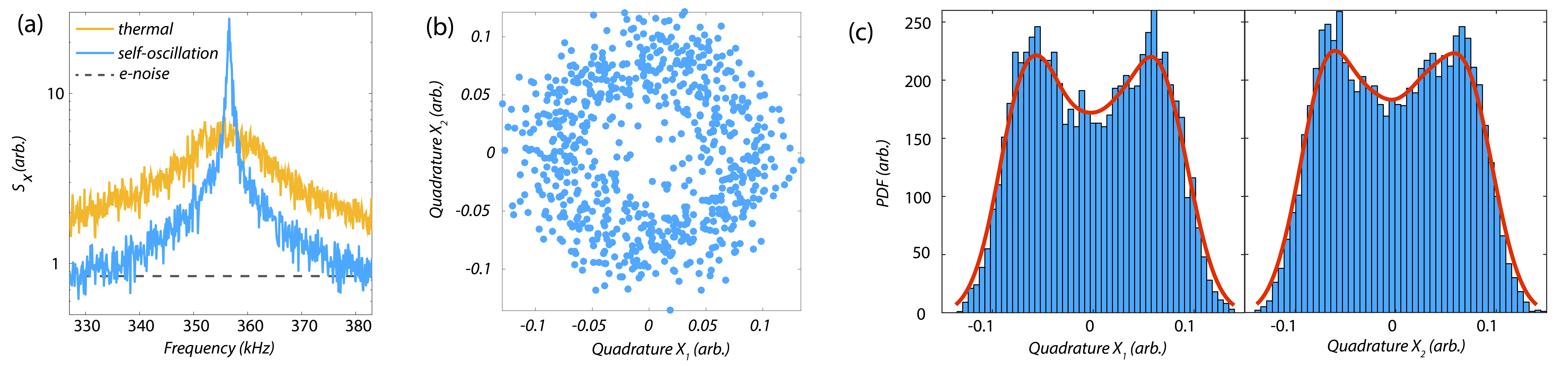}
	\caption{\textbf{E-beam induced Dynamical Backaction (device D3).} (a) Mechanical spectra for a damped thermal state (broad) and a self-oscillating state (narrow). The electronic noise background level (grey). (b) A hole in the phase-space of the associated quadratures is observed, indicating a self-oscillating mechanical state. (c) Histograms of the motion quadratures presenting non-Gaussian statistics. }
	\label{fig:figure5}
\end{figure}


\end{document}